\documentclass[twocolumn,floatfix,showpacs,showkeys,preprintnumbers,nofootinbib,superscriptaddress]{revtex4}
\usepackage{amsmath}
\usepackage{graphicx}
\usepackage{textcomp}
\usepackage{amsfonts}
\usepackage{amssymb}
\usepackage{bm}
\usepackage{multirow}
\usepackage{booktabs}
\usepackage{subfigure}
\usepackage{float}

\usepackage[outdir=./]{epstopdf}

\begin{document}

\title{Investigations on the charmless decays of $X(3872)$ in intermediate meson loops model}

\author{Yan Wang}  
\affiliation{College of Physics and Engineering, Qufu Normal
University, Qufu 273165, China}
\author{Qi Wu} 
\affiliation{School of Physics and Center of High Energy Physics, Peking University, Beijing 100871, China}
\author{Gang Li}\email{gli@qfnu.edu.cn}
\affiliation{College of Physics and Engineering, Qufu Normal
University, Qufu 273165, China}
\author{Wen-Hua Qin}  \email{qwh@qfnu.edu.cn}
\affiliation{College of Physics and Engineering, Qufu Normal
University, Qufu 273165, China}
\author{Xiao-Hai Liu}\email{xiaohai.liu@tju.edu.cn}
\affiliation{Department of Physics, School of Science, Tianjin
University, Tianjin 300350, China}
\author{Chun-Sheng An}\email{ancs@swu.edu.cn}
\affiliation{School of Physical Science and Technology, Southwest University, Chongqing 400715, China}
\author{Ju-Jun Xie}\email{xiejujun@impcas.ac.cn}
\affiliation{Institute of Modern Physics, Chinese Academy of
Sciences, Lanzhou 730000, China} \affiliation{School of Nuclear
Science and Technology, University of Chinese Academy of Sciences,
Beijing 101408, China} \affiliation{School of Physics and
Microelectronics, Zhengzhou University, Zhengzhou, Henan 450001,
China }

\begin{abstract}

The charmless decay processes of $X(3872)$ provide us a good platform to study the nature and the decay mechanism of $X(3872)$. Based on a molecular nature of $X(3872)$ as a $\bar{D}D^*$ bound state, we have investigated the charmless decays $X(3872) \to VV$ and $VP$ via intermediate $D^*{\bar D} +c.c.$ meson loops, where $V$ and $P$ stand for light vector and pseudoscalar mesons, respectively. We discuss three cases, i.e., pure neutral components ($\theta=0$), isospin singlet ($\theta=\pi/4$) and neutral components dominant ($\theta = \pi/6$), where $\theta$ is a phase angle describing the proportion of neutral and charged constituents. The proportion of neutral and charged constituent have an influence on the decay widths of $X(3872) \to VV$ and $VP$. With the coupling constant of $X(3872)$ to the $\bar{D}D^*$ channel obtained under the molecule ansatz of $X(3872)$ resonance, the predicted decay widths of $X(3872)\rightarrow VV$ are about tens of keVs, while the decay width can reach a few hundreds of keVs for $X(3872)\to VP$. The dependence of these ratios between different decay modes of $X(3872)\to VV$ and $X(3872)\to VP$ to the mixing angle $\theta$ is also investigated. It is expected that the theoretical calculations here can be tested by future experiments.

\end{abstract}

\maketitle

\section{Introduction}  \label{sec:introduction}

In 2003, the $X(3872)$ state was first observed by the Belle
Collaboration in the $J/\psi \pi^+ \pi^-$ invariant mass spectrum of the $B\to KX(3872)\to K \pi^+\pi^-
J/\psi$ decay~\cite{Choi:2003ue}. Then, it was confirmed in the $J/\psi
\pi^+ \pi^-$ channel from $p\bar{p}$ collisions by CDF and D0
Collaborations~\cite{Acosta:2003zx,Abazov:2004kp}, and $e^+ e^-$
collisions by BABAR
Collaboration~\cite{Aubert:2004ns,Aubert:2008gu}. Its quantum
numbers was determined to be $I^G (J^{PC})=0^+ (1^{++})$ by LHCb
Collaboration~\cite{Aaij:2013zoa}. There are two salient feature of
$X(3872)$, one is that it has very narrow width $(\Gamma_X<1.2~
\mathrm{MeV})$, the other one is that its mass is extremely closing
to the mass threshold of $D^0 \bar{D}^{\ast0}$ channel.

The interpretation of the nature of $X(3872)$ is still an open
question. Since its quantum numbers are $J^{PC}=1^{++}$ and its mass
is very close to the $D^0 \bar{D}^{\ast0}$ threshold, one naturel
explanation is that it is a $D\bar{D}^\ast$ hadronic molecule as
discussed in
Refs.~\cite{Close:2003sg,Pakvasa:2003ea,Swanson:2004pp,Swanson:2003tb,Tornqvist:2004qy,Voloshin:2003nt,Wong:2003xk,AlFiky:2005jd,Braaten:2006sy,Fleming:2007rp,
Ding:2009vj,Dong:2009yp,Lee:2009hy,Lee:2011rka,Liu:2008tn,Zhang:2009vs,Gamermann:2009uq,Mehen:2011ds,Nieves:2011vw,Nieves:2012tt,Li:2012cs,Sun:2012sy,Guo:2013sya,
Guo:2014hqa,He:2014nya,Zhao:2014gqa,Guo:2014taa,Braaten:2003he}. In
general, a hadronic molecule can couple to other components which
have the same quantum numbers. For instance, the possibility of a
charmonium $c\bar{c}$ excited state admixture was investigated in
Refs.~\cite{Dong:2008gb,Suzuki:2005ha}. It was also pointed out that
the $D^\pm D^{\ast\mp}$ and $D^+_s D^{\ast-}_s$ components are
necessary to explain the branching ratio of $X(3872)$ to $J/\psi
\rho$ and $J/\psi \omega$~\cite{Gamermann:2007fi,Gamermann:2009fv,Aceti:2012cb}. On
the other hand, the $X(3872)$ is also considered as a tetraquark
state~\cite{Maiani:2004vq,Maiani:2005pe,Maiani:2007vr,Terasaki:2007uv}.
However, searching for the charged partners of $X(3872)$ shows negative
results~\cite{Aubert:2004zr}. Besides, the $X(3872)$ was also viewed
as a conventional charmonium
state~\cite{Barnes:2003vb,Eichten:2004uh,Barnes:2005pb}.

In Ref.~\cite{Liu:2006df}, the isospin violating decay process of
$X(3872)\rightarrow J/\psi \rho$ was estimated using final state
interactions (FSI) by consider intermediate $D\bar{D}^\ast$ meson
loop, where it was found that the contribution from FSI is tiny. The
radiative decays $X(3872)\rightarrow \gamma\psi/\psi^\prime$
were investigated in Refs.~\cite{Dong:2009uf,Dong:2009yp,Guo:2014taa}, and the results
support the molecular picture of $X(3872)$. While in
Refs.~\cite{Dubynskiy:2007tj,Fleming:2008yn,Mehen:2015efa}, the
pionic transition from $X(3872)$ to $\chi_{cJ}$ was studied. In
Ref.~\cite{Dubynskiy:2007tj} it was concluded that these decay rates
exhibit significantly different patterns depending on a pure
charmonium or a multi-quark structure of $X(3872)$.

All these above theoretical studies of $X(3872)$ focus on its
charmful decay modes. To better understand the nature of $X(3872)$,
the study of its other decay modes is needed. For example, the
charmless decays can also provide us a good platform to further
study the nature of $X(3872)$. In this work, under the molecule ansatz of the $X(3872)$, which is a bound state of $\bar{D}D^*$, we will investigate the
charmless decays of $X(3872) \to VV$ and $VP$ ($V$ and $P$ stand for the
vector meson and pseudoscalar meson) via intermediate charmed meson
loops in an effective Lagrangian approach.

This article is organized as follows. In Sec.~\ref{Sec:charmless}, based on a molecular nature of $X(3872)$ as a $\bar{D}D^*$ bound state, we present the related decay amplitudes obtained with the effective Lagrangians constructed in the heavy quark limit and chiral
symmetry. In Sec.~\ref{Sec:Num}, we show our numerical results and discussions, and last section is devoted to a short summary.

\section{Theoretical framework} \label{Sec:charmless}

\subsection{Coupling constant and decay diagrams}

For a state slightly below an S-wave two-hadron threshold, the effective coupling constant of this state to the two-body channel, $g_{\rm eff}$, is related to the probability of finding the two-hadron component in the physical wave function of the bound state, $c_i^2$, and the binding energy $\epsilon=m_1+m_2-M$~\cite{Weinberg:1965zz,Baru:2003qq,Guo:2013zbw}
\begin{eqnarray}\label{eq:coupling-X}
g_{{\rm eff}}^2 = 16\pi c_i^2 (m_1+ m_2)^2 \sqrt{\frac
{2\epsilon}{\mu}} , \label{eq:lag}
\end{eqnarray}
where $\mu=m_1 m_2/(m_1 + m_2)$ is the reduced mass of $m_1$ and $m_2$.

Assuming that the $X(3872)$ is a $S$-wave molecular state with quantum numbers $J^{PC}=1^{++}$ given by the superposition of $D^0{\bar D}^{*0}$ and $D^\pm D^{\ast\mp}$ hadronic configurations as
\begin{eqnarray}
|X(3872)\rangle &=&  \frac {\cos\theta} {\sqrt{2}} |D^{*0}{\bar D}^0 +D^{0}{\bar D}^{*0}\rangle  \nonumber \\
&& + \frac{\sin\theta} {\sqrt{2}} |D^{*+}D^- + D^{-}D^{*+}\rangle, \label{eq:wavef}
\end{eqnarray}
where $\theta$ is a phase angle describing the proportion of neutral and charged constituents. For example, $\theta=0$ stands for $X(3872)$ as a pure ${\bar D}^{*0}D^0/{\bar D}^{0}D^{*0}$, while $\theta=\pi/4$ and $\theta=-\pi/4$ correspond to the isospin singlet and isospin triplet states, respectively.
Then, one can parameterize the coupling of $X(3872)$ to the charmed
mesons in terms of the following Lagrangian:

\begin{eqnarray}
{\cal L}_{X(3872)} &=&  \frac {g_n} {\sqrt{2}} X_\mu^\dag(D^{*0\mu}{\bar D}^0 +D^{0}{\bar D}^{*0\mu})  \nonumber \\
&& + \frac{g_c} {\sqrt{2}} X_\mu^\dag (D^{*+\mu}D^- + D^{+}D^{*-\mu}), \label{eq:wavef}
\end{eqnarray}
where $g_n$ and $g_c$ are the coupling constants of $X(3872)$ with its neutral and charged components, respectively.

Using the masses of the $X(3872)$ and the charmed mesons as in Refs.~\cite{Zyla:2020zbs,Guo:2013zbw}, we obtain the mass difference between the
$X(3872)$ and the ${\bar D}^{*0}D^0/{\bar D}^0D^{*0}$ (neutral) and $D^{*-}D^+/D^{*+}D^-$ threshold to be $0.16$ MeV and $8.21$ MeV, respectively. Assuming that $X(3872)$ is a pure $D^0{\bar D}^{*0}$ or $D^{\pm}D^{*\mp}$ molecule, we obtain
\begin{eqnarray}
|g_{{\rm eff}}^n|&=&3.70 ~{\rm GeV}, ~~~ {\rm with}~c^2_{D^0{\bar D}^{*0}}=1,  \\
|g_{{\rm eff}}^c|&=&9.91 ~{\rm GeV}, ~~~ {\rm with}~c^2_{D^{\pm}D^{*\mp}}=1.\label{eq:values}
\end{eqnarray}
As a result, the coupling constants appearing in Eq. (\ref{eq:wavef}) are as follows~\footnote{These coupling constants are assumed to be real.},
\begin{eqnarray}\label{eq:coupling}
g_n=|g_{{\rm eff}}^n|\cos \theta, ~~~  g_c=|g_{{\rm eff}}^c|\sin \theta \, .
\end{eqnarray}

\begin{figure}[htbp]
\centering
\includegraphics[width=0.45\textwidth]{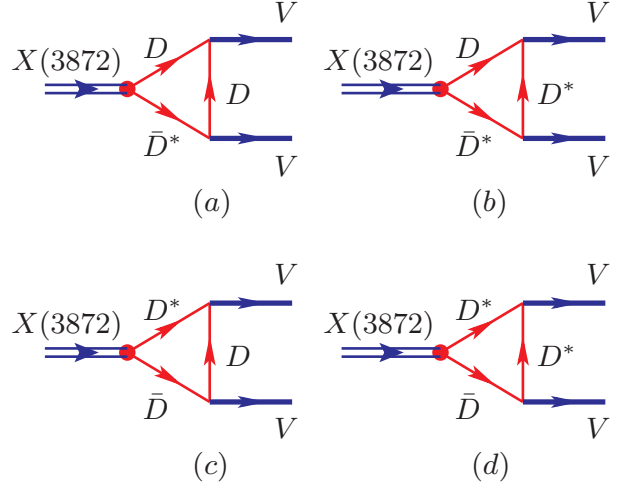}
\caption{Diagrams contributing to the charmless decay $X(3872) \to
VV$ with $D{\bar D}^*+c.c.$ as intermediate states.}
\label{fig:feyn-VV}
\end{figure}

\begin{figure}[htbp]
\centering
\includegraphics[width=0.45\textwidth]{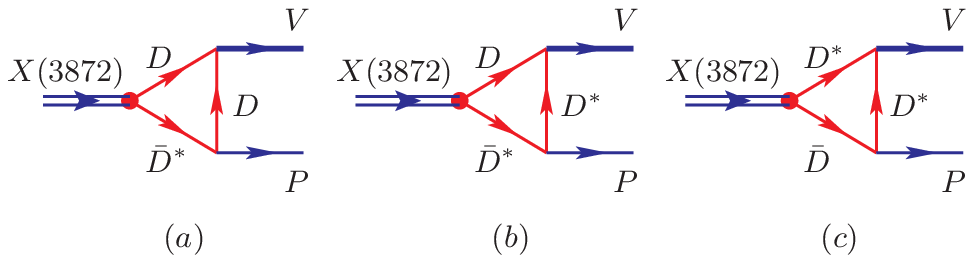}
\caption{Diagrams contributing to the charmless decay $X(3872) \to
VP$ with $D{\bar D}^*+c.c.$ as intermediate states.}
\label{fig:feyn-VP}
\end{figure}

With the above $\bar{D}D^*$ molecular picture for $X(3872)$, these
$X(3872) \to VV$ and $VP$ decays can proceed via $X(3872) \to
\bar{D}D^* \to VV$ or $VP$ through triangle loop diagrams, which are
shown in Figs.~\ref{fig:feyn-VV} and \ref{fig:feyn-VP},
respectively. In this mechanism, $X(3872)$ goes into $\bar{D}D^*$ at a first step, then $\bar{D}$ and $D^*$ are converged to $VV$ or $VP$ in the final state by exchanging a charmed meson. Note that, in Figs.~\ref{fig:feyn-VV} and \ref{fig:feyn-VP}, we have considered only the leading contributions as discussed in Refs.~\cite{Lipkin:1986bi,Lipkin:1988tg,Lipkin:1986av}.

\subsection{The interaction Lagrangians and decay amplitudes}

The Lagrangians relevant to the light vector and pseudoscalar mesons can be constructed based on the heavy quark limit and chiral
symmetry,
 \begin{eqnarray}
 {\cal L} &=& -ig_{{\cal D}^{\ast }{\cal D}
{\mathcal P}}\left( {\cal D}^i \partial^\mu {\mathcal P}_{ij} {\cal D}_\mu^{\ast
j\dagger }-{\cal D}_\mu^{\ast i}\partial^\mu {\mathcal P}_{ij} {\cal D}^{j \dag}\right) \nonumber \\
&& +\frac{1}{2}g_{{\cal D}^\ast D^\ast {\mathcal P}}\varepsilon _{\mu
\nu \alpha \beta }{\cal D}_i^{\ast \mu }\partial^\nu {\mathcal P}^{ij}  {\overset{
\leftrightarrow }{\partial }}{\!^{\alpha }} {\cal D}_j^{\ast \beta\dag } \nonumber \\
&& - ig_{\mathcal{D}\mathcal{D}\mathcal{V}} \mathcal{D}_i^\dagger {\stackrel{\leftrightarrow}{\partial}}{\!_\mu} \mathcal{D}^j(\mathcal{V}^\mu)^i_j \nonumber \\
&& -2f_{\mathcal{D}^*\mathcal{D}\mathcal{V}} \epsilon_{\mu\nu\alpha\beta}
(\partial^\mu \mathcal{V}^\nu)^i_j
(\mathcal{D}_i^\dagger{\stackrel{\leftrightarrow}{\partial}}{\!^\alpha} \mathcal{D}^{*\beta j}-\mathcal{D}_i^{*\beta\dagger}{\stackrel{\leftrightarrow}{\partial}}{\!^\alpha} {\cal D}^j) \nonumber
\\
&&+ ig_{\mathcal{D}^*\mathcal{D}^*\mathcal{V}} \mathcal{D}^{*\nu\dagger}_i {\stackrel{\leftrightarrow}{\partial}}{\!_\mu} \mathcal{D}^{*j}_\nu(\mathcal{V}^\mu)^i_j \nonumber \\
&& +4if_{\mathcal{D}^*\mathcal{D}^*\mathcal{V}} \mathcal{D}^{*\dagger}_{i\mu}(\partial^\mu \mathcal{V}^\nu-\partial^\nu
\mathcal{V}^\mu)^i_j \mathcal{D}^{*j}_\nu +{\rm H.c.} , \label{eq:light-meson}
 \label{eq:LDDV}
 \end{eqnarray}
with the convention $\varepsilon_{0123}=1$, where $\mathcal P$ and ${\mathcal V}_\mu$
are $3\times 3$ matrices for the octet pseudoscalar and nonet vector
mesons, respectively,
 \begin{eqnarray}
 \setlength{\abovedisplayskip}{3pt}
 \setlength{\belowdisplayskip}{3pt}
   \mathcal{P} &= &\left(
\begin{array} {ccc}
\frac {\pi^0} {\sqrt {2}} +\frac {\eta} {\sqrt{6}} & \pi^+ & K^+ \\
\pi^- & -\frac {\pi^0} {\sqrt {2}} +\frac { \eta} {\sqrt{6}} & K^0 \\
K^-& {\bar K}^0 & -\sqrt{\frac{2}{3}}\eta \\
\end{array}\right), \\
\mathcal{V} &=& \left(\begin{array}{ccc}\frac{\rho^0} {\sqrt {2}}+\frac {\omega} {\sqrt {2}}&\rho^+ & K^{*+} \\
\rho^- & -\frac {\rho^0} {\sqrt {2}} + \frac {\omega} {\sqrt {2}} & K^{*0} \\
K^{*-}& {\bar K}^{*0} & \phi \\
\end{array}\right) \, .
\end{eqnarray}

In the heavy quark and chiral limits, the couplings of the charmed meson  to the light vector mesons have the
relationship~\cite{Casalbuoni:1996pg,Cheng:2004ru},
\begin{eqnarray}
\setlength{\abovedisplayskip}{3pt}
 \setlength{\belowdisplayskip}{3pt}
g_{{\cal D}{\cal D}V} &=& g_{{\cal D}^*{\cal D}^*V}=\frac{\beta g_V}{\sqrt{2}} , \\
f_{{\cal D}^*{\cal D}V} &=& \frac{ f_{{\cal D}^*{\cal D}^*V}}{m_{{\cal D}^*}}=\frac{\lambda g_V}{\sqrt{2}} \, , \\
g_{\mathcal{D}^{*} \mathcal{D} \mathcal{P}} &=& \frac{2 g}{f_{\pi}}
\sqrt{m_{\mathcal{D}} m_{\mathcal{D}^{*}}}, \\
g_{\mathcal{D}^{*} \mathcal{D}^{*} \mathcal{P}} &=& \frac{g_{{\cal
D}^{*} {\mathcal D} {\mathcal {P}}}}{\sqrt{m_{\mathcal{D}}
m_{\mathcal{D}^{*}}}} \, .
\end{eqnarray}
In this work, we take parameters $\beta=0.9$,  $\lambda = 0.56 \,{\rm GeV}^{-1} $, $g=0.59$, and $g_V = {m_\rho /f_\pi}$ with $f_\pi = 132$ MeV, as used in previous works~\cite{Casalbuoni:1996pg,Isola:2003fh}.

Then one can easily write the explicit transition amplitudes for
$X(3872)(p_1)\to [D^{(*)}(q_1) {\bar D}^{(*)}(q_3)] D^{(*)}(q_2) \to
V_1(p_2)V_2(p_3)$ shown in Fig.~\ref{fig:feyn-VV} as follows:

\begin{eqnarray}
\mathcal{M}_a&=& \int\frac{d^4q_2}{(2\pi)^4}[g_{\rm eff}\epsilon_{1\alpha}][g_{DDV}(q_1-q_2)_\mu\epsilon_2^{*\mu}]\nonumber\\
&&\times[2f_{D^*DV}\epsilon_{\kappa\lambda\rho\sigma}ip_3^{\kappa}\epsilon_3^{*\lambda}(q_2+q_3)^\rho]\frac{1}{q_1^2-m_1^2}\nonumber\\
&&\times\frac{1}{q_2^2-m_2^2}\frac{(g^{\alpha\sigma} - {q_3^\alpha
q_3^\sigma}/{m_3^2})}{q_3^2-m_3^2}{\cal F}(q^2),
\end{eqnarray}
\begin{eqnarray}
\mathcal{M}_b&=& \int\frac{d^4q_2}{(2\pi)^4}[g_{\rm eff}\epsilon_{1\alpha}][2f_{D^*DV}\epsilon_{\mu\nu\xi\phi}p_2^\mu\epsilon_2^{*\nu}(q_1-q_2)^\xi]\nonumber\\
&&\times[g_{D^*D^*V}(q_3+q_2)^\rho g_{\lambda\sigma}\epsilon^*_{3\rho} + 4f_{D^*D^*V}(p_{3\lambda}g_\sigma^\rho\nonumber\\
&&-p_{3\sigma}g_\lambda^\rho)\epsilon^*_{3\rho}] \frac{i}{q_1^2-m_1^2}\frac{(g^{\phi\sigma} - {q_2^\phi q_2^\sigma}/{m_2^2})}{q_2^2-m_2^2}\nonumber\\
&&\times\frac{(g^{\alpha\lambda} - {q_3^\alpha
q_3^\lambda}/{m_3^2})}{q_3^2-m_3^2}{\cal F}(q^2),
\end{eqnarray}
\begin{eqnarray}
\mathcal{M}_c&=& \int\frac{d^4q_2}{(2\pi)^4}[g_{\rm eff}\epsilon_{1\alpha}][-2f_{D^*DV}\epsilon_{\mu\nu\xi\phi}p_2^\mu\epsilon_2^{*\nu}(q_1-q_2)^\xi]\nonumber\\
&&\times[(g_{DDV}(q_3+q_2)_\kappa\epsilon_3^{*\kappa}]\frac{(g^{\alpha\phi} - {q_1^\alpha q_1^\phi}/{m_1^2})}{q_1^2-m_1^2}\nonumber\\
&& \times \frac{i}{q_2^2-m_2^2}\frac{1}{q_3^2-m_3^2}{\cal F}(q^2),
\end{eqnarray}
\begin{eqnarray}
\mathcal{M}_d&=& \int\frac{d^4q_2}{(2\pi)^4}[g_{\rm eff}\epsilon_{1\alpha}][g_{D^*D^*V}(q_1-q_2)^\xi g_{\nu\phi}\epsilon_{2\xi}^*\nonumber\\
&& - 4f_{D^*D^*V}(p_{2\nu}g_\phi^\xi - p_{2\phi}g_\nu^\xi)\epsilon_{2\xi}^*]\nonumber\\
&&\times[-2f_{D^*DV}\epsilon_{\kappa\lambda\rho\sigma}p^\kappa_3 \epsilon_3^{*\lambda}(q_2+q_3)^\rho]\nonumber\\
&&\times\frac{(g^{\alpha\phi}- {q_1^\alpha q_1^\phi}/{m_1^2})}{q_1^2-m_1^2}\frac{(g^{\nu\sigma} - {q_2^\nu q_2^\sigma}/{m_2^2})}{q_2^2-m_2^2}\nonumber\\
&&\times\frac{i}{q_3^2-m_3^2}{\cal F}(q^2) \, ,\label{eq:X-VV}
\end{eqnarray}
where $p_1$ ($\varepsilon_1$), $p_2$ ($\varepsilon_2$) and $p_3$
($\varepsilon_3$) are the four-momenta (polarization vector) of the
initial state $X(3872)$, final state $V_1$ and $V_2$,
respectively. $q_1$, $q_2$ and $q_3$ are the four-momenta of the up,
right and down charmed mesons in the triangle loop, respectively.

The explicit transition amplitudes for $X(3872)(p_1)\to
[D^{(*)}(q_1) {\bar D}^{(*)}(q_3)] D^{(*)}(q_2) \to V(p_2)P(p_3)$
shown in Fig.~\ref{fig:feyn-VP} are as follows:
\begin{eqnarray}
\mathcal{M}_a &=& \int\frac{d^4q_2}{(2\pi)^4}[g_{\rm eff}\epsilon_{1\alpha}][-g_{DDV}(q_1-q_2)_\mu\epsilon_2^{*\mu}]\nonumber\\
&&\times[g_{D^*DP}p_3^\kappa]\frac{1}{q_1^2-m_1^2}\frac{i}{q_2^2-m_2^2}
\nonumber \\
&& \times \frac{(g_\kappa^\alpha - {q_3^\alpha
q_{3\kappa}}/{m_3^2})}{q_3^2-m_3^2}{\cal F}(q^2), \\
\mathcal{M}_b&=&\int\frac{d^4q_2}{(2\pi)^4}[g_{\rm eff}\epsilon_{1\alpha}][2f_{D^*DV}\epsilon_{\mu\nu\xi\phi}p_2^\mu\epsilon_2^{*\nu}(q_1-q_2)^\xi]\nonumber\\
&&\times[\frac{1}{2}g_{D^*D^*P}\epsilon_{\kappa\lambda\rho\sigma} p_3^\lambda (q_3+q_2)^\rho]\frac{-i}{q_1^2-m_1^2}\nonumber\\
&&\times\frac{(g^{\phi\kappa} - {q_2^\phi
q_2^\kappa}/{m_2^2})}{q_2^2-m_2^2}\frac{(g^{\alpha\sigma} -
{q_3^\alpha q_3^\sigma}/{m_3^2})}{q_3^2-m_3^2}{\cal F}(q^2),
\end{eqnarray}
\begin{eqnarray}
\mathcal{M}_c&=& \int\frac{d^4q_2}{(2\pi)^4}[g_{\rm eff}\epsilon_{1\alpha}][g_{D^*D^*V}(q_1-q_2)^\xi g_{\nu\phi}\epsilon_{2\xi}^*\nonumber\\
&& - 4f_{D^*D^*V}(p_{2\nu}g_\phi^\xi - p_{2\phi}g_\nu^\xi)\epsilon_{2\xi}^*][g_{D^*DP}p_3^\kappa]\nonumber\\
&&\times\frac{(g^{\alpha\phi} - {q_1^\alpha q_1^\phi}/{m_1^2})}{q_1^2-m_1^2}\frac{(g_\kappa^\nu - {q_2^\nu q_{2\kappa}}/{m_2^2})}{q_2^2-m_2^2}\nonumber\\
&&\times\frac{i}{q_3^2-m_3^2}{\cal F}(q^2),
\end{eqnarray}
where $\mathcal{F} (q^2)$ is the form factor introduced to depict
the off-shell effects of the exchanged mesons as well as the
structure effects of the involved mesons. The form factor $\mathcal{F} (q^2)$ is parameterized as
\begin{eqnarray}
	\mathcal{F}\left(q^{2}\right)=
	\left(\frac{m^{2} - \Lambda^{2}}{q^{2} - \Lambda^{2}} \right)^n,
	\label{Eq:FFsl}
\end{eqnarray}
normalized to unity at $q^2 = m^2$~\cite{Cheng:2004ru}, where $m$ and $q$ are mass and momenta of the exchanged mesons. The cutoff $\Lambda$ can be further reparameterized as $\Lambda=m_{D^{(\ast)}}+\alpha\Lambda_{\rm QCD} $ with $\Lambda_{\rm QCD}=0.22 \ {\rm GeV}$. The model parameter $\alpha$ is usually expected to be of order of
unity~\cite{Cheng:2004ru,Tornqvist:1993vu,Tornqvist:1993ng,Locher:1993cc,Li:1996yn},
but its concrete value cannot be estimated by the first principle. In practice, the value of $\alpha$ is usually determined by comparing theoretical estimates with the corresponding experimental measurements. However, no charmless decay mode of $X(3872)$ is known so far. For the rescattering processes studied in this work, it is found that the monopole form ($n=1$) or dipole form ($n=2$) for $\mathcal{F} (q^2)$ is utilized, the numerical results are much sensitive to the values of parameter $\alpha$, and we have to use a very small value, otherwise, these partial decay widths will be very large, even more than the total width of $X(3872)$. In order to avoid too large dependence of the parameter $\alpha$, we take $n=3$ in the numerical calculations.

\section{Numerical Results and discussions}  \label{Sec:Num}

\begin{figure}[htbp]
\centering
\includegraphics[width=0.4\textwidth]{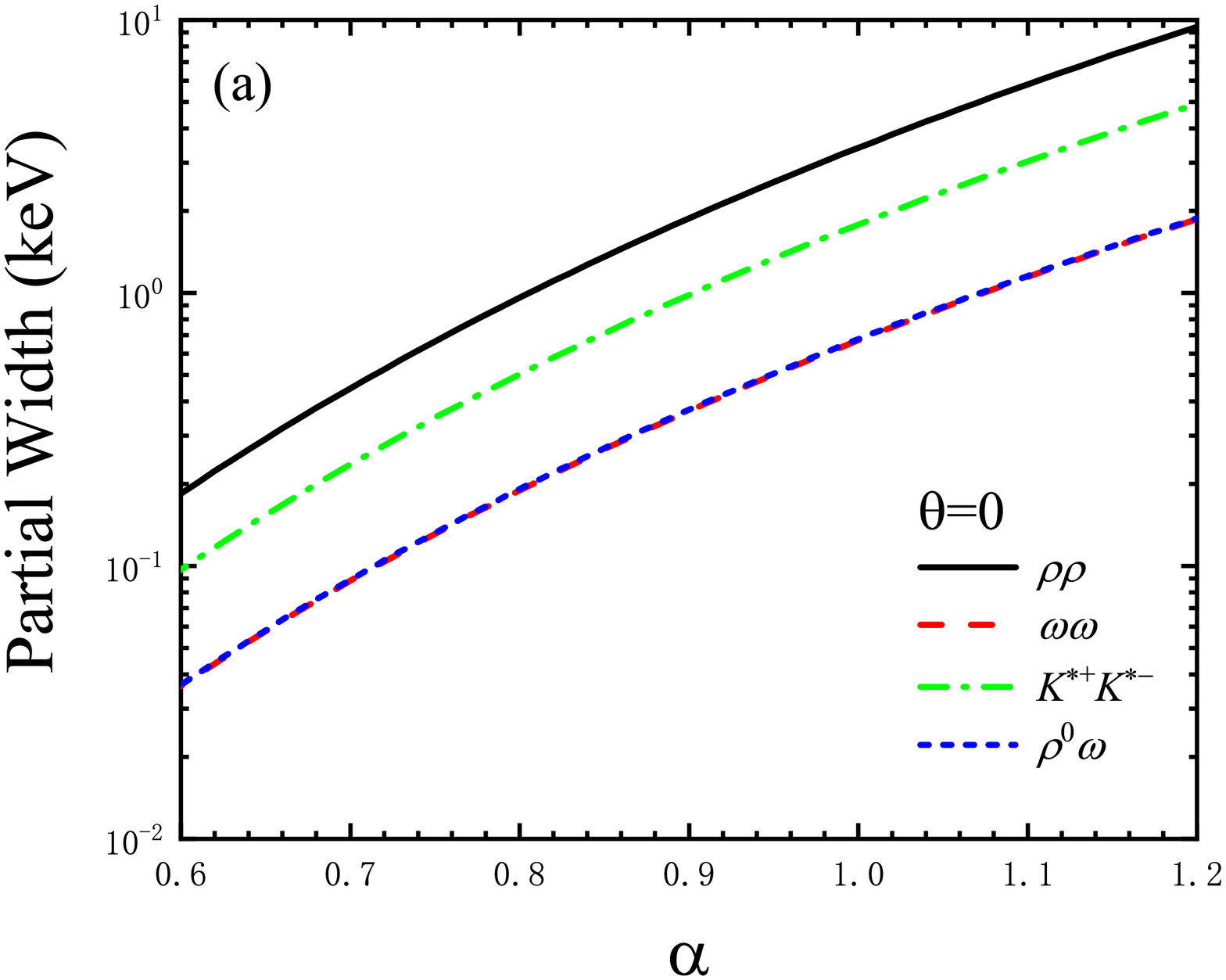}
\includegraphics[width=0.4\textwidth]{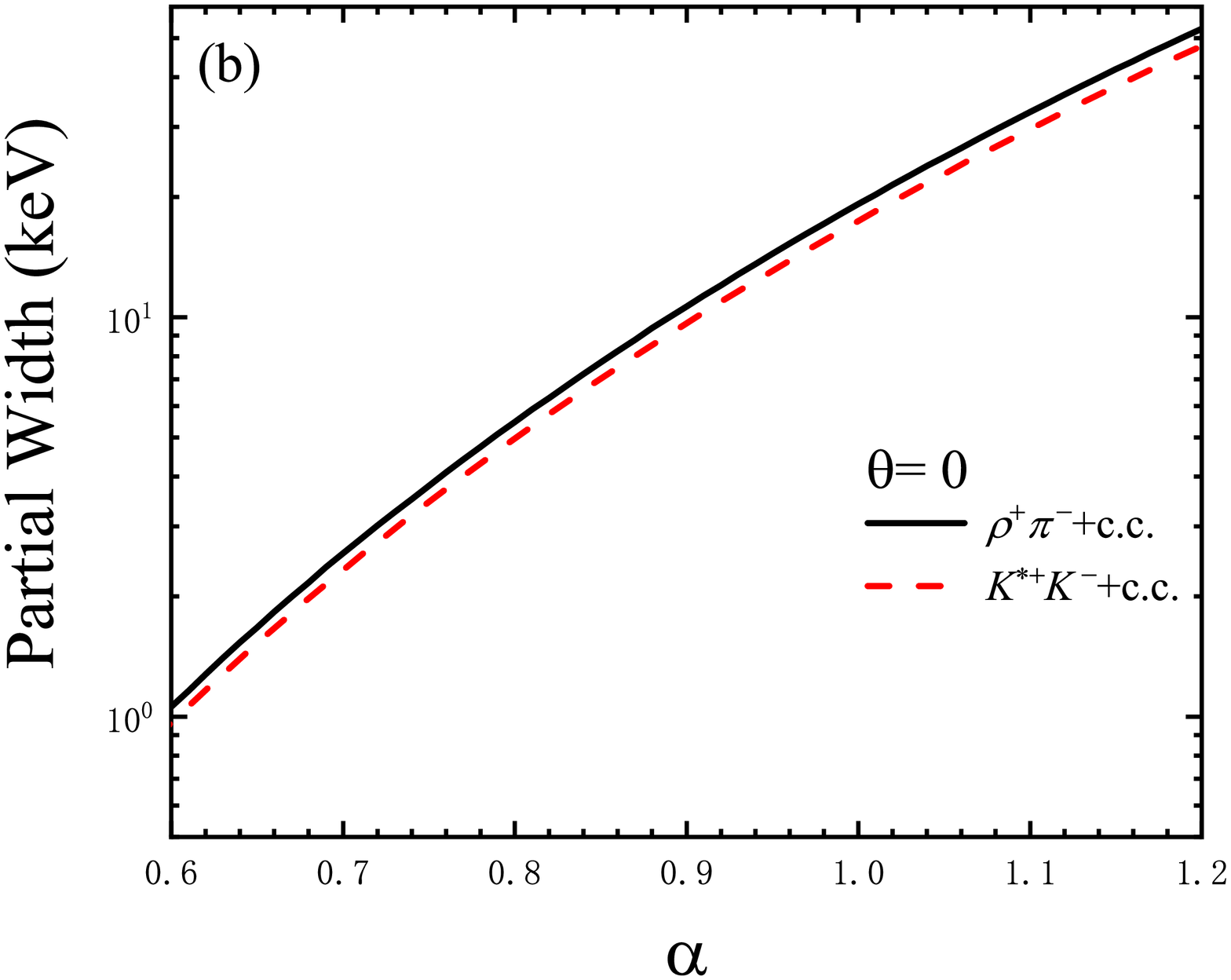}
\caption{The $\alpha$-dependence of decay widths (in unit of keV) of $X(3872)\rightarrow VV$ and $X(3872)\rightarrow VP$ with $\theta=0$.}
\label{fig:x3872-0}
\end{figure}

\begin{figure}[htbp]
\includegraphics[width=0.4\textwidth]{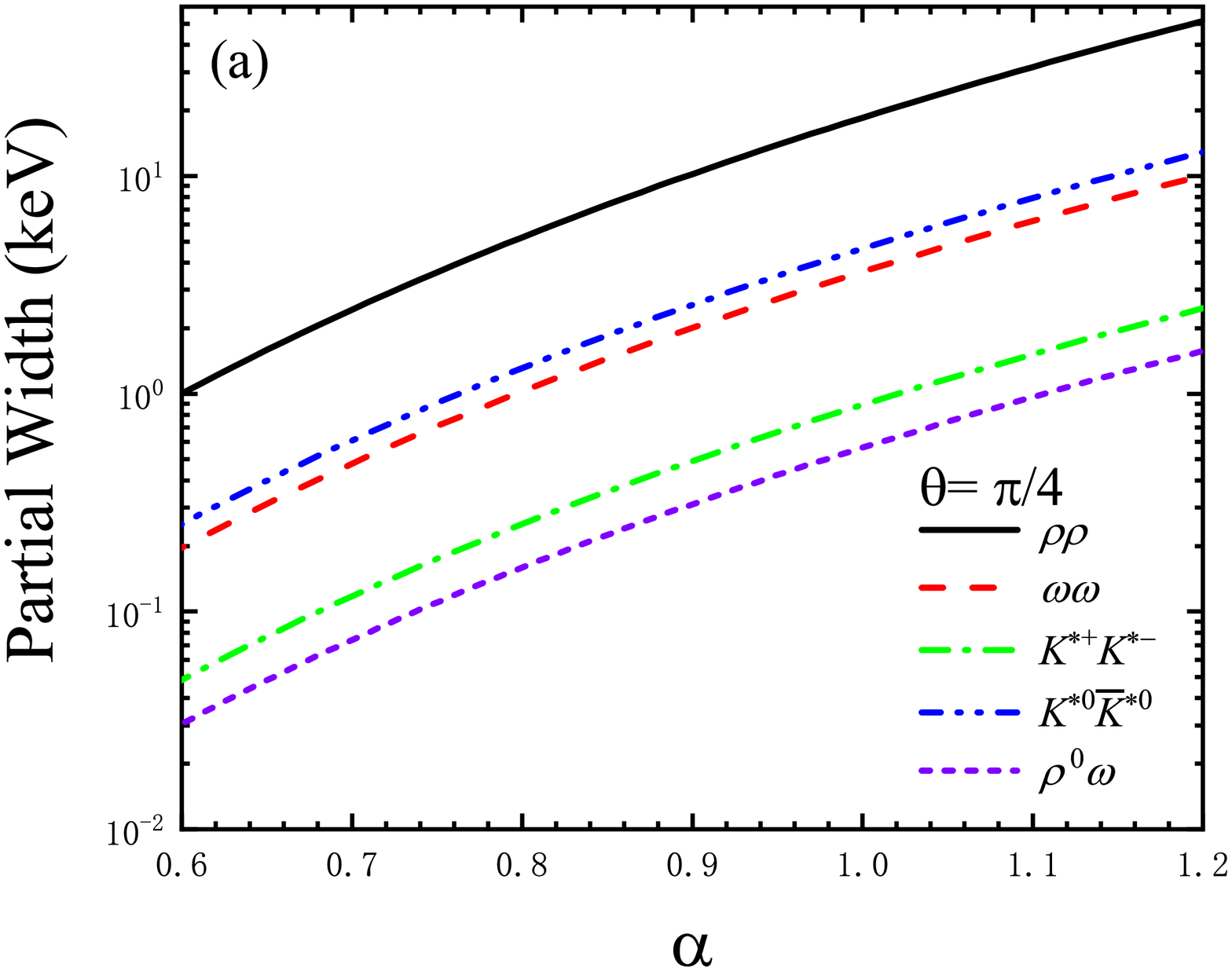}
\includegraphics[width=0.4\textwidth]{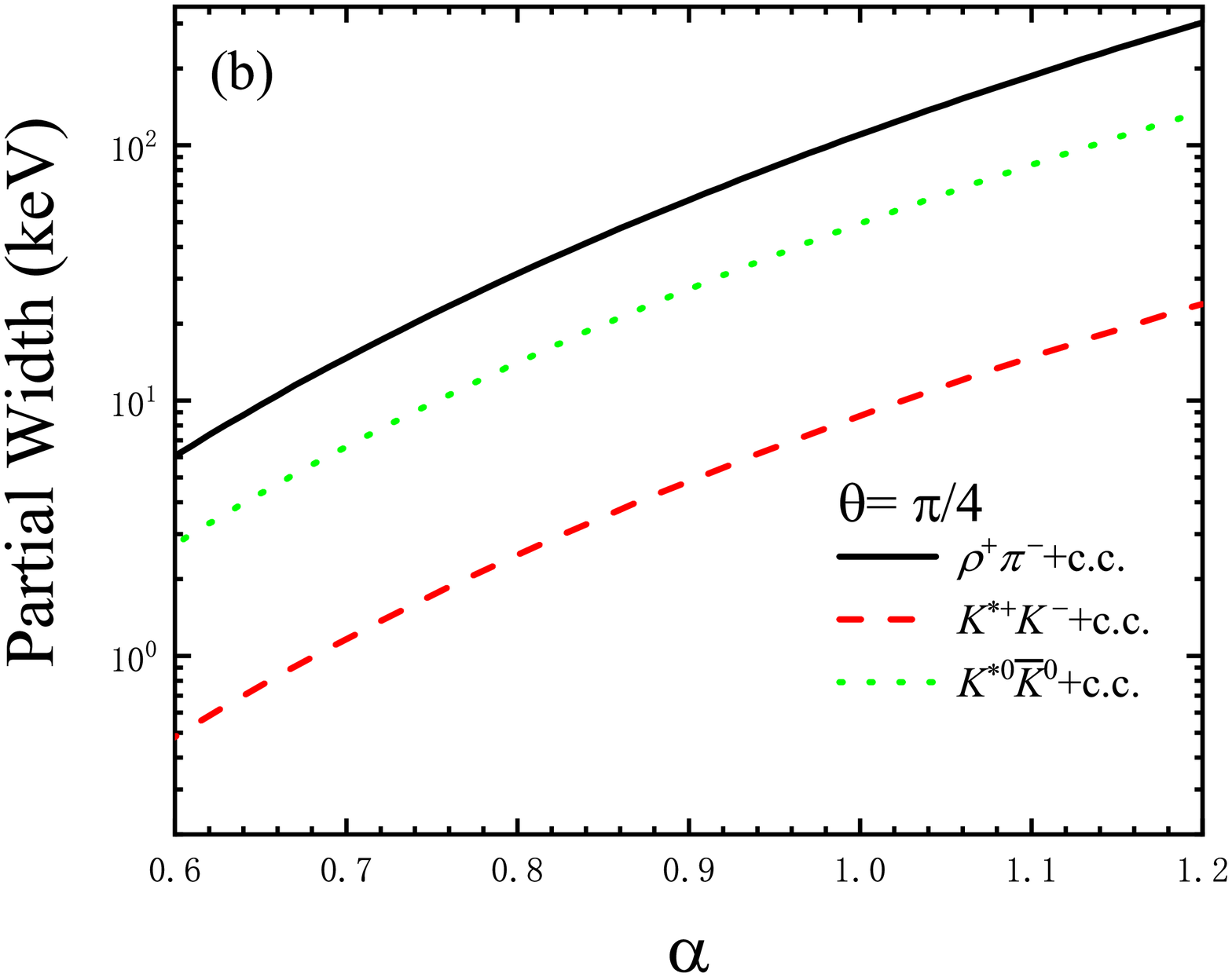}
\caption{The $\alpha$-dependence of decay widths (in unit of keV) of $X(3872)\rightarrow VV$ and $X(3872)\rightarrow VP$ with $\theta=\pi/4$}
\label{fig:x3872-45}
\end{figure}

\begin{figure}[htbp]
\includegraphics[width=0.4\textwidth]{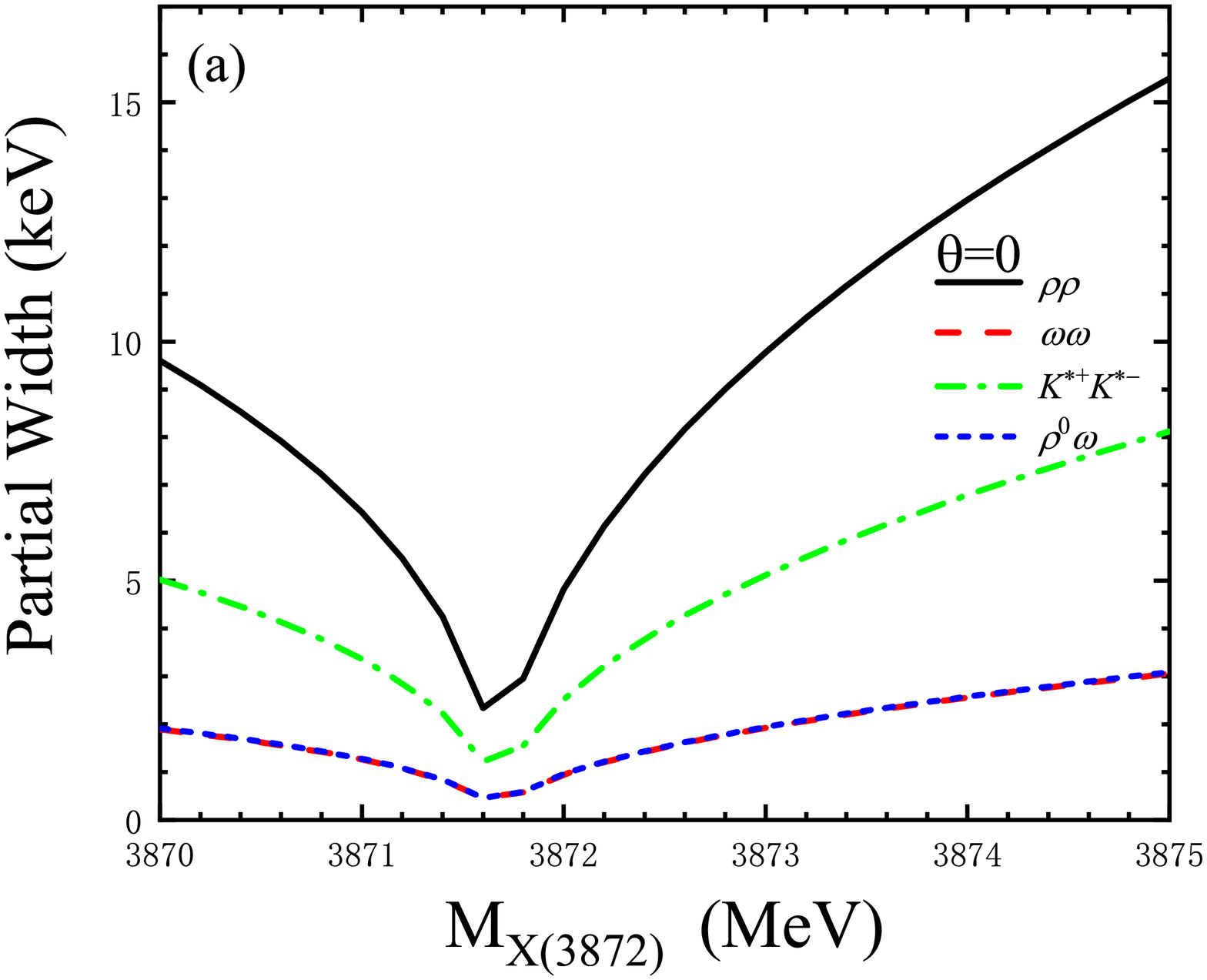}
\includegraphics[width=0.4\textwidth]{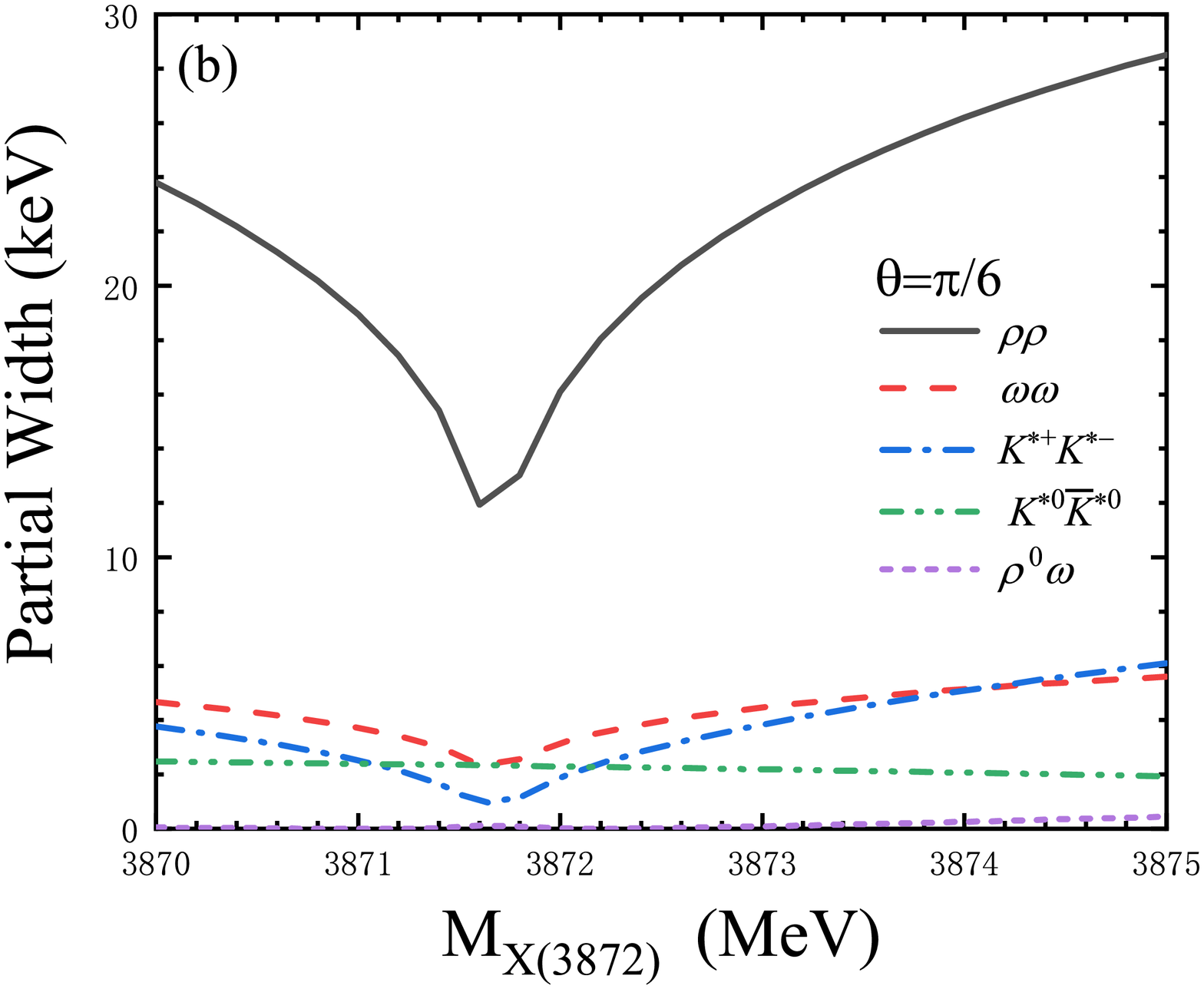}
\includegraphics[width=0.4\textwidth]{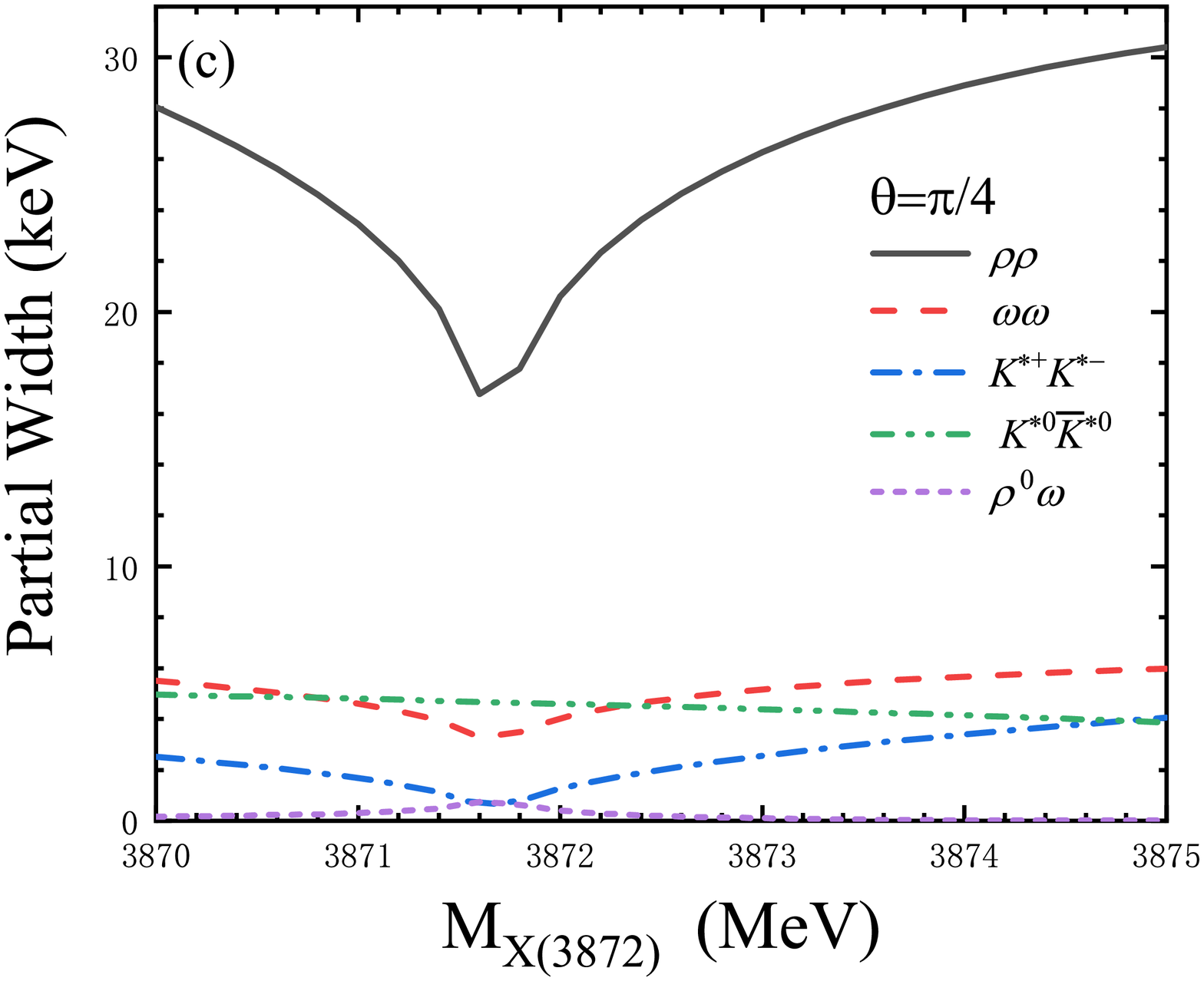}
\caption{ The $M_{X(3872)}$-dependence of the decay widths (in unit of keV) of $X(3872)\rightarrow VV$ with $\alpha=1.0$.}
\label{fig:vvbe}
\end{figure}

\begin{figure}[htbp]
\includegraphics[width=0.4\textwidth]{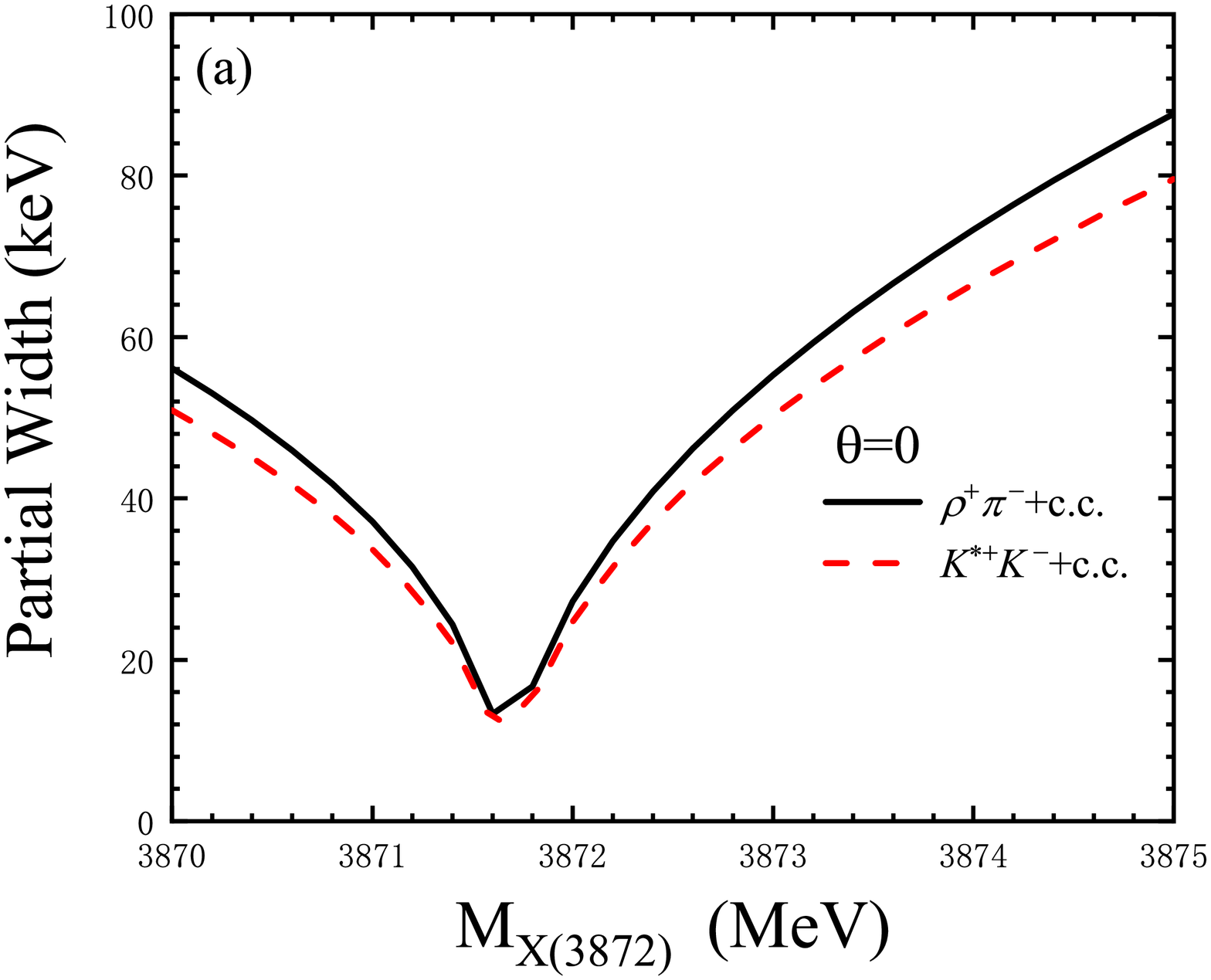}
\includegraphics[width=0.4\textwidth]{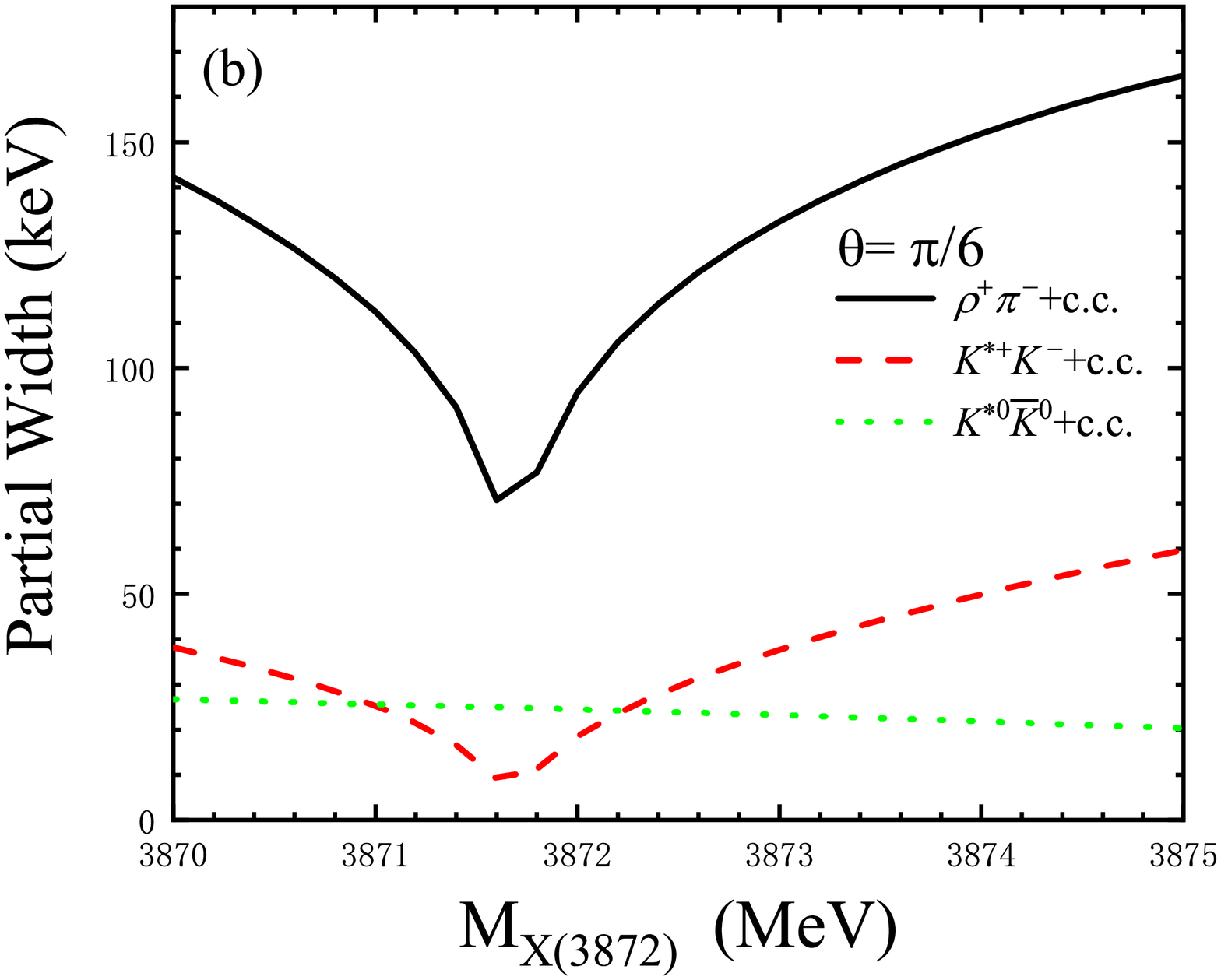}
\includegraphics[width=0.4\textwidth]{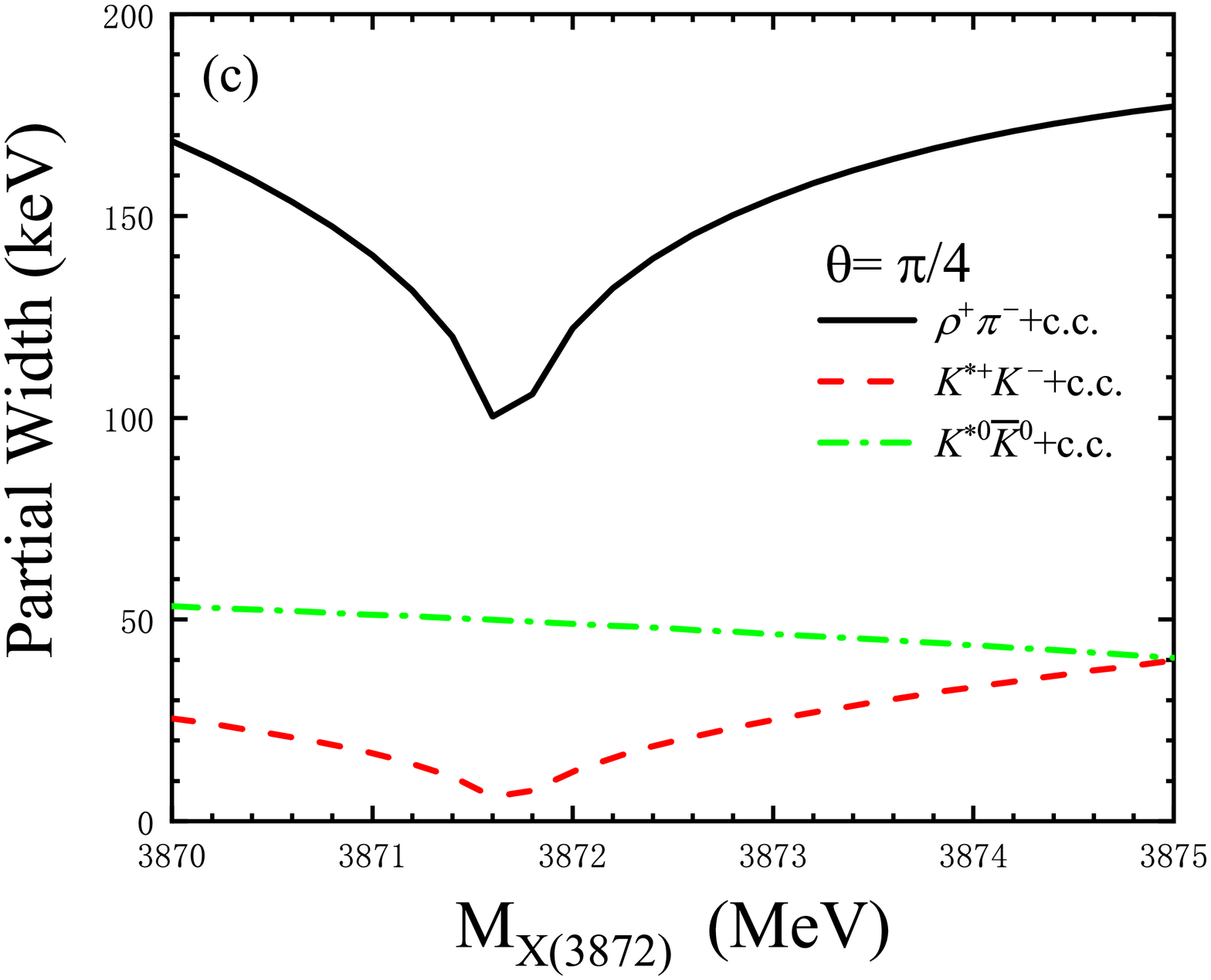}
\caption{The same as Fig~\ref{fig:vvbe} but for $X(3872)\rightarrow VP$.}
\label{fig:vpbe}
\end{figure}

In this section, we will mainly discuss three cases where $\theta$ is $0$, $\pi/6$ and $\pi/4$. When $\theta=0$, it indicates that $X(3872)$ is a pure bound state with only neutral component. When $\theta=\pi/4$, the proportions of the neutral and charged components are the same. There are both neutral and charged components at $\theta=\pi/6$, but the proportion of the neutral component is dominant.

In Fig.~\ref{fig:x3872-0}, we plot the $\alpha$-dependence of the partial
decay widths of $X(3872)\rightarrow VV$ and $X(3872)\rightarrow VP$ with $\theta=0$, respectively. In the range of $\alpha=0.6\sim 1.2$, the predicted partial decay
widths of $X(3872)\rightarrow VV$ are about a few KeV, while the partial decay widths can reach a few tens of KeV for $X(3872) \to VP$. Since the $X(3872)\to K^{\ast 0}\bar{K}^{\ast 0}$ transition proceeds via
$[D^+D^{*-}]D^{(\ast)}_s$ intermediate mesons, while the
$X(3872)\to K^{\ast+}K^{\ast-}$ transition proceeds via $[D^0\bar{D}^{*0}]D^{(\ast)}_s$ intermediate mesons. So in the case of $\theta=0$, there is no neutral $K^{\ast0}\bar{K}^{\ast0}$ channel as shown in Fig.~\ref{fig:x3872-0} (a). The same reason for $X(3872)\rightarrow K^{\ast0}\bar{K}^{0}$ in Fig.~\ref{fig:x3872-0} (b). From Fig.~\ref{fig:x3872-0} (a), one can see that the partial decay width of $X(3872)\to\rho\rho$ is larger than those of $X(3872) \to K^{\ast+}K^{\ast-}$ and $\omega\omega$
decay modes. This is because both the charged $\rho^+ \rho^-$ and neutron $\rho^0 \rho^0$ channels contribute to the $\rho\rho$ channel. While for the $X(3872) \to \rho^0 \rho^0$ decay, its partial decay width is almost equal to the decay of $ X(3872) \to \omega\omega$. In addition, for the $X(3872) \to K^{\ast+}K^{\ast-}$ decays, there are only contributions from the exchanging of charged charm mesons.  In the case of $\theta=0$, only neutral charmed
meson loops contribute to the isospin-violating channel $X(3872) \to \rho^0\omega$. As a result, the obtained decay widths are almost the same as that of the channel $X(3872) \to \omega\omega$.

In Fig.~\ref{fig:x3872-45}, we plot the $\alpha$-dependence of the partial
decay widths of $X(3872)\rightarrow VV$ and $X(3872)\rightarrow VP$ with $\theta=\pi/4$. In the range of $\alpha=0.6\sim 1.2$, the predicted partial decay
widths of $X(3872)\rightarrow VV$ are about a few tens of KeV, while the partial decay widths can reach several hundred KeV for $X(3872) \to VP$. The behavior is similar to that of Fig.~\ref{fig:x3872-0}. Since the case of $\theta=\pi/4$ corresponds to equal neutral and charged components in $X(3872)$, so the channels $X(3872) \to K^{\ast+}K^{\ast-}$ and $X(3872) \to K^{\ast 0}\bar{K}^{\ast0}$ have non-zero decay widths. The $X(3872)\to K^{\ast 0}\bar{K}^{\ast 0}$ transition proceed via
$[D^+D^{*-}]D^{(\ast)}_s$ intermediate mesons, while the
$X(3872)\to K^{\ast+}K^{\ast-}$ transition proceed via $[D^0\bar{D}^{*0}]D^{(\ast)}_s$ intermediate mesons. The mass of
$X(3872)$ is much closer the mass threshold of $D^0\bar{D}^{*0}$ than $D^+D^{*-}$, so the threshold effects of $X(3872)\to K^{\ast+}K^{\ast-}$ will be larger than that of $X(3872)\to K^{\ast 0}\bar{K}^{\ast 0}$. However,  the couplings constant values obtained from Eq.~(\ref{eq:coupling}) have the relation $g_n \textless g_c$. Thus with the same value of $\alpha$, the obtained partial decay width of $X(3872)\rightarrow K^{\ast0}\bar{K}^{\ast0}$ is about several times larger than that of $X(3872)\rightarrow K^{\ast+}K^{\ast-}$. However, for the $X(3872) \to \rho\rho$ decay, there are contributions from the exchanging both charged charm mesons and neutral charm mesons, and these two contributions give the instructive interference of the decay amplitudes. A similar situation occurs in
$X(3872)\rightarrow VP$ as shown in Fig.~\ref{fig:x3872-0} (b). A similar situation occurs in
$X(3872)\rightarrow VP$ as shown in Fig.~\ref{fig:x3872-45} (b). In the case of $\theta=\pi/4$, the charged and neutral charmed
meson loops should cancel out exactly in the isospin
symmetry limit for the isospin-violating channel $X(3872) \to \rho^0\omega$.  In other words, the mass difference between the $u$ and $d$ quark will lead to $m_{D^{(*)\pm}} \neq m_{D^{(*)0}}$
due to the isospin symmetry breaking. As a result, the
charged and neutral charmed meson loops cannot completely cancel out, and the residue part will contribute to
the isospin-violating amplitudes. The partial widths of
the isospin-violating channel $X(3872) \to \omega\rho^0$ as shown in Fig.~\ref{fig:x3872-45} (a) are
suppressed.

Using the center value of the total decay width of $X(3872)$ that was reported recently by the LHCb Collaboration~\cite{Aaij:2020qga,Aaij:2020xjx}, we obtain the branching ratios for $X(3872)\rightarrow VV$ and $VP$  in the cases of $\theta=0$, $\pi/6$ and $\pi/4$ , respectively. We take the range of $\alpha$ as $0.6\sim1.2$, then the numerical results are shown in the Table.~\ref{tab:a}. Our theoretical numerical results show that with the increase of $\theta$, the partial decay widths of $K^{\ast+}K^{\ast-}$ and $K^{\ast+}K^{-}+c.c.$ channels decrease. Because there is only neutral charmed mesons loops in $X(3872)\rightarrow K^{\ast+}K^{\ast-}$ and $X(3872)\rightarrow K^{\ast+}K^{-}$. And also the $X(3872)$ coupling constant to the neutral channel $g_n$ is proportional to $\cos\theta$.

\begin{table*}[ht]
\begin{center}
\caption{The branching ratios for $X(3872)\rightarrow VV$ and $X(3872)\rightarrow VP$ with different $\theta$ values. The $\alpha$ range is taken to be $0.6\sim1.2$ here.}\label{tab:a}
\begin{tabular}{ccccccccc}
\hline \hline
Final states    & $\theta=0$  & $\theta=\pi/6$  & $\theta=\pi/4$   \\
\hline
$\rho\rho$                     &$(0.15-7.86)\times 10^{-3}$    &$(0.06-3.20)\times 10^{-2}$   &$(0.83-4.29)\times 10^{-2}$    \\
$K^{\ast+}K^{\ast-}$           &$(0.08-4.11)\times 10^{-3}$    &$(0.06-3.08)\times 10^{-3}$   &$(0.04-2.05)\times 10^{-3}$    \\
$K^{\ast0}\bar{K}^{\ast 0}$         &        $ -- $                 &$(0.11-5.36)\times 10^{-3}$   &$(0.02-1.07)\times 10^{-2}$   \\
$\omega\omega$                 &$(0.03-1.55)\times 10^{-3}$   &$(0.12-6.28)\times 10^{-3}$   &$(0.16-8.41)\times 10^{-3}$    \\
$\rho^0\omega$                 &$(0.03-1.56)\times 10^{-3}$   &$(0.02-1.25)\times 10^{-4}$      &$(0.03-1.31)\times 10^{-3}$       \\
$\rho^\pm \pi^\mp$             &$(0.09-4.40)\times 10^{-2}$    &$(0.004-1.87)\times 10^{-1}$   &$(0.05-2.53)\times 10^{-1}$    \\
$K^{\ast+}K^{-}+c.c.$          &$(0.08-3.99)\times 10^{-2}$    &$(0.06-2.99)\times 10^{-2}$   &$(0.04-1.99)\times 10^{-2}$  \\
$K^{\ast0}\bar{K}^{0}+c.c.$    &         $--$                  &$(0.11-5.66)\times 10^{-2}$   &$(0.02-1.13)\times 10^{-1}$ \\
\hline \hline
\end{tabular}
\end{center}
\end{table*}

In Fig.~\ref{fig:vvbe}, we present the partial decay widths of the $X(3872)\rightarrow VV$ in terms of the mass of $X(3872)$, where we have fixed the value of $\alpha$ as $1.0$. The coupling constant of $X(3872)$ in
Eq.~(\ref{eq:coupling-X}) and the threshold effects can simultaneously influence the the mass of $X(3872)$ dependence of the decay widths. Generally speaking, with increasing the mass difference between $X(3872)$ and $D^{*0} {\bar D}^0$ mesons, i.e., increasing the
binding energy, the coupling strength of $X(3872)$
increases, and the threshold effects decrease. Both the coupling
strength of $X(3872)$ and the threshold effects vary quickly in the
small binding energy region and slowly in the large binding energy
region. As a result, the behavior of the partial widths is
relatively sensitive at small binding energy, while it becomes smooth
at large binding energy. The single-cusp structure locates at the the thresholds of the $D^{*0} {\bar D}^0$ mesons for most of the decay channels except for $K^{*0}\bar{K}^{*0}$ channel. This is because the $X(3872)\to K^{\ast 0}\bar{K}^{\ast 0}$ transition proceed via
$[D^+D^{*-}]D^{(\ast)}_s$ intermediate mesons. A similar behavior of partial widths occur in $X(3872)\rightarrow VP$ as shown in the Fig.~\ref{fig:vpbe}.

It would be interesting to further clarify the uncertainties arising from the introduction of form factors by studying the $\alpha$ dependence of the ratios between different partial decay widths.
For the decays $X(3872) \to VV$, we define the
following ratios to the partial decay widths of $X(3872) \to \omega \omega$


\begin{eqnarray}
R_1 &=& \frac {\Gamma(X(3872) \to \omega \rho^0)} {\Gamma (X(3872) \to \omega \omega)} \, , \nonumber \\
R_2 &=& \frac {\Gamma(X(3872) \to \rho \rho)} {\Gamma (X(3872) \to \omega \omega)} \, , \nonumber \\
R_3 &=& \frac {\Gamma(X(3872) \to K^{\ast +} K^{\ast -} )} {\Gamma (X(3872) \to \omega \omega)} \, , \nonumber \\
R_4 &=& \frac {\Gamma(X(3872) \to K^{\ast0}\bar{K}^{\ast 0})} {\Gamma (X(3872) \to \omega \omega)} \, .
\label{eq:ratio-VV}
\end{eqnarray}

For the decays of $X(3872) \to VP$, the following ratios are defined:
\begin{eqnarray}
r_1 &=& \frac {\Gamma(X(3872) \to K^{\ast +} K^- +c.c.)} {\Gamma (X(3872) \to \rho \pi)} \, , \nonumber \\
r_2 &=& \frac {\Gamma(X(3872) \to K^{\ast0}\bar{K}^0 +c.c.)} {\Gamma (X(3872) \to \rho \pi)} \, .
\label{eq:ratio-VP}
\end{eqnarray}

\begin{figure}[htbp]
\includegraphics[width=0.4\textwidth]{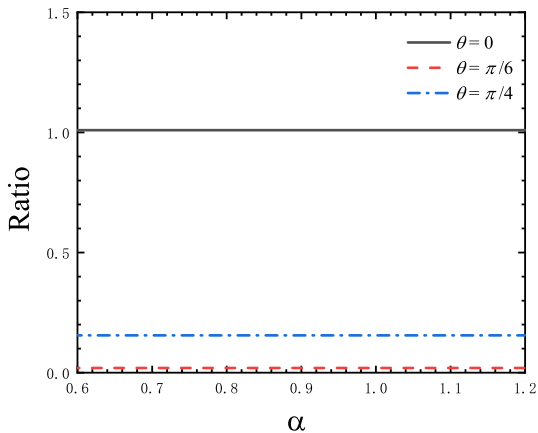}
\caption{The $\alpha$-dependence of the ratio $R_1$ defined in Eq.~(\ref{eq:ratio-VV}).}
\label{fig:ratio}
\end{figure}

The ratios $R_1$ in terms of $\alpha$ are plotted in Fig.~\ref{fig:ratio}. The results of Fig.~\ref{fig:ratio} show that the ratios are completely insensitive to this dependence. This stabilities of the ratios in terms of $\alpha$ indicate a reasonably controlled cutoff for each channel by the form factor to some extent. On the other hand, one can see that, in Fig.~\ref{fig:ratio}, there are extremely strong dependence of the ratio on the isospin mixing angle, $\theta$, which is of more fundamental significance than the parameter $\alpha$. This stability stimulate us to study the mixing angle $\theta$ dependence.

\begin{figure*}[htbp]
\centering
\hspace{-2.5cm}
\includegraphics[scale=0.35]{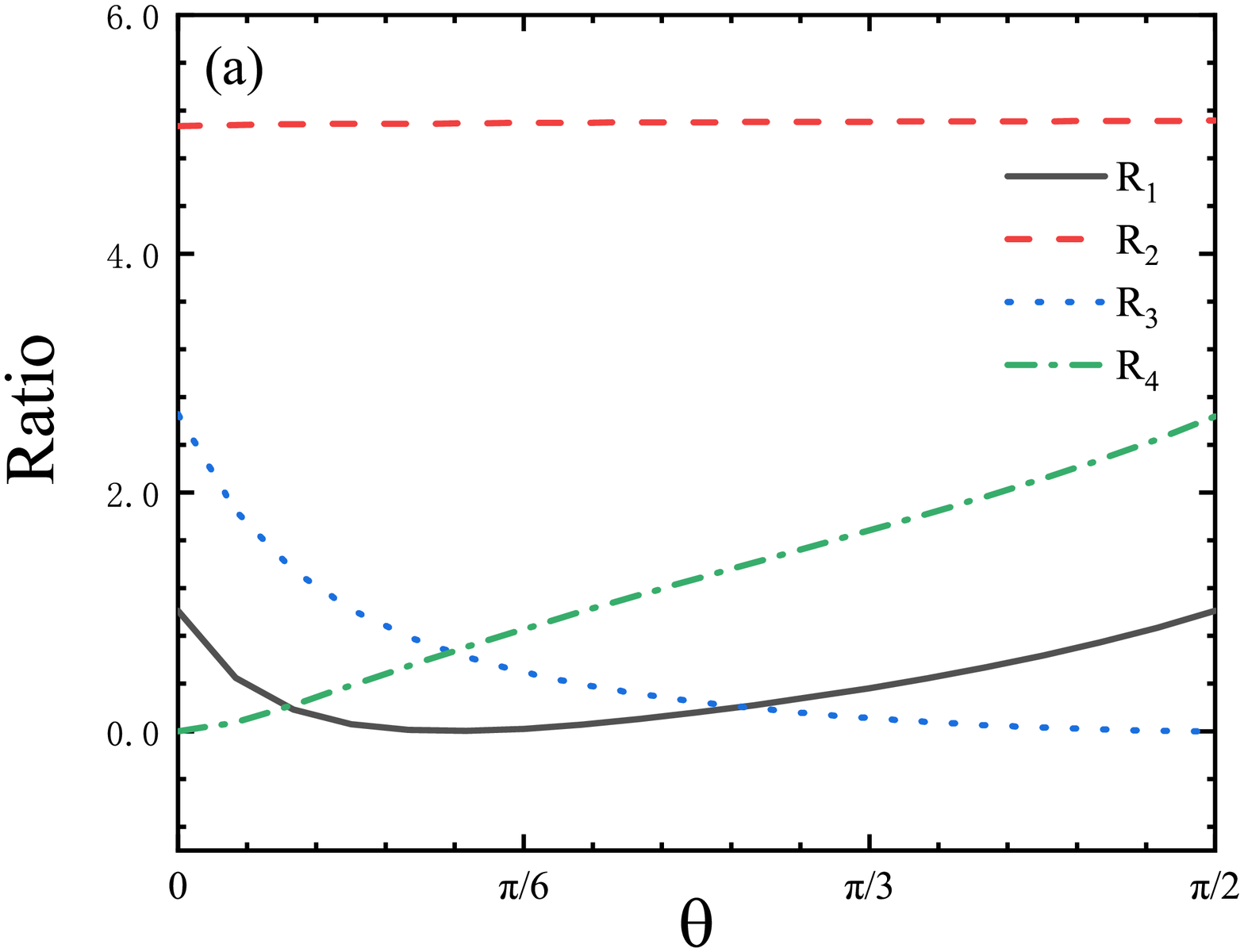}
\includegraphics[scale=0.35]{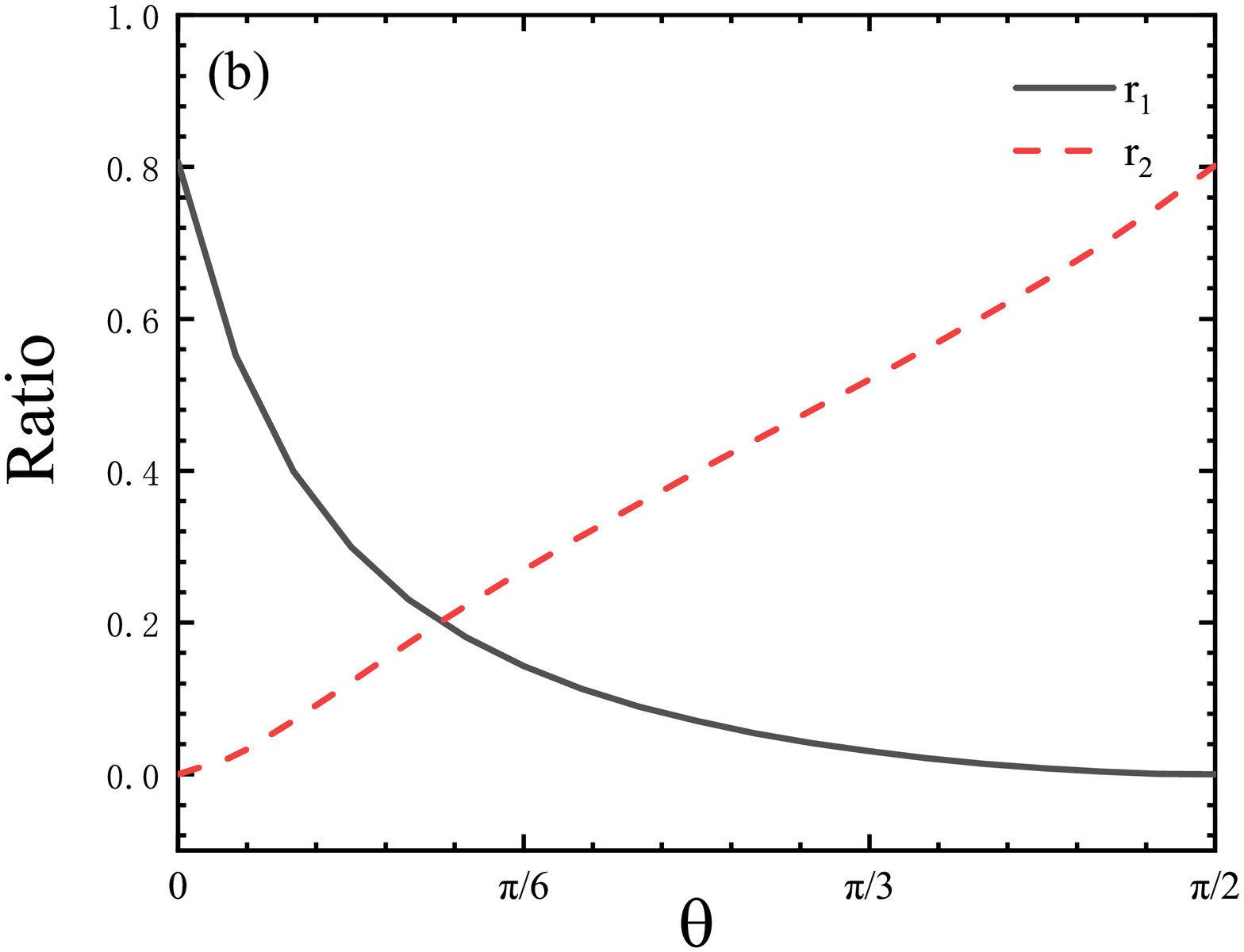}
\caption{(a). The ratio $R_i$ $(i=1,2,3,4)$ defined in Eq.~(\ref{eq:ratio-VV}) as a function of the mixing angle $\theta$ with $\alpha=1.0$. (b). The ratio $r_i$ $(i=1,2)$ defined in Eq.~(\ref{eq:ratio-VP}) as a function of the mixing angle $\theta$ with $\alpha=1.0$.}
\label{fig:ratio-theta}
\end{figure*}

Next, we turn to the dependence of these ratios defined in Eqs.~\eqref{eq:ratio-VV} and \eqref{eq:ratio-VP} to the mixing angle $\theta$ with a fixed $\alpha$. In Fig.~\ref{fig:ratio-theta}, we present the theoretical results of the ratio $R_i$ $(i=1,2,3,4)$ defined in Eq.~(\ref{eq:ratio-VV}) and $r_i$ $(i=1,2)$ defined in Eq.~(\ref{eq:ratio-VP}) as a function of the mixing angle $\theta$ with a fixed value $\alpha=1.0$. It is interesting to note that the results of the ratio $R_2 = \frac {\Gamma(X(3872) \to \rho \rho)} {\Gamma (X(3872) \to \omega \omega)}$ are not dependent on the value of $\theta$. These ratios shown in Fig.~\ref{fig:ratio-theta} may be tested by the future experimental measurements and can be used to determine the value of the mixing angle.

\section{Summary}  \label{Sec:Summary}

Based on a molecular nature of $X(3872)$,  we have investigated the charmless decays of
$X(3872) \to VV$ and $VP$. For $X(3872)$, we considered three cases, i.e., pure neutral components ($\theta=0$), isospin singlet ($\theta=\pi/4$) and neutral components dominant ($\theta = \pi/6$), where $\theta$ is a phase angle describing the proportion of neutral and charged constituents. We explore the rescattering mechanism
within the effective Lagrangian based on the heavy quark symmetry and
chiral symmetry. We can see that although the decay widths increase with the increase of $\alpha$ when we fix the phase angle $\theta$, our theoretical results show that the cutoff parameter $\alpha$ dependence of the partial widths is not drastically sensitive, which
indicates the dominant mechanism driven by the intermediate meson
loops with a fairly good control of the ultraviolet contributions. When $X(3872)$ is a pure neutral bound state, the predicted partial decay
widths of $X(3872)\rightarrow VV$ are about a few keV, while the partial decay widths can reach a few tens of keV for $X(3872) \to VP$. When there are both neutral and charged components in $X(3872)$, the predicted decay widths of $X(3872)\rightarrow VV$ are about tens of keV. while the decay widths can reach a few hundreds of keV for $X(3872)\rightarrow VP$.

Moreover, the dependence of these ratios between different charmless decay modes of $X(3872)$ to the charged and neutral
mixing angle for the $X(3872)$ in the molecular picture is also investigated, which may be tested by future experiments and can be used to determine the value of the mixing angle.

\section* {Acknowledgements}

We thank the anonymous referee for very constructive comments on the manuscript. This work is supported by the National Natural Science Foundation of China, under Grants Nos. 12075133, 11835015, 11975165, 12075288, 11735003, 11675131, and 11961141012. It is also partly supported by Taishan Scholar Project of Shandong Province (Grant No. tsqn202103062), the Higher Educational Youth Innovation Science and Technology Program Shandong Province (Grant No. 2020KJJ004), and the Youth Innovation Promotion
Association CAS.

\end{document}